\documentclass{elsart4}

 \usepackage{graphicx}

\usepackage{amssymb}

\begin{document}

\begin{frontmatter}



\title{Doping effects in low dimensional antiferromagnets}


\author[KL,CH]{Sebastian Eggert}
\author[CH]{and Fabrizio Anfuso}

\address[KL]{Physics Dept., Univ. of Kaiserslautern, 67663 Kaiserslautern, Germany}
\address[CH]{Institute of Theoretical Physics, Chalmers University of Technology,
S-412 96 G\"oteborg, Sweden}

\begin{abstract}
The study of impurities in low dimensional antiferromagnets has been a very active field
in magnetism ever since the discovery of high temperature superconductivity.
One of the most dramatic effects  is the appearance of large Knight shifts in a
long range around non-magnetic impurities in an antiferromagnetic background.
The dependence of the Knight shifts on distance and temperature visualizes the
correlations in the system.
In this work we consider
the Knight shifts around a single vacancy in the one and two dimensional Heisenberg model.
\end{abstract}

\end{frontmatter}

Doping and impurity effects in low-dimensional quantum antiferromagnets
remain of strong interest in the condensed
matter physics community, spurred by high-$T_c$ superconductivity
and other exotic effects in those systems.  
One of the most basic questions to ask is how an antiferromagnetic systems responds to 
a uniform applied magnetic field and how this response is changed in the presence of 
impurities.  This question will be answered in detail for a two-dimensional (2D) spin-1/2
Heisenberg model.  The results will then be compared to the one-dimensional spin-chain model 
which does not order at low temperatures.

\begin{figure}
\includegraphics[width=0.45\textwidth]{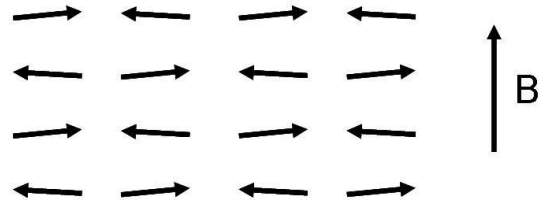}
\caption{The effect of a magnetic field $B$ on a generic antiferromagnet.}
\label{AF-B}
\end{figure}

\begin{figure}
\includegraphics[width=0.45\textwidth]{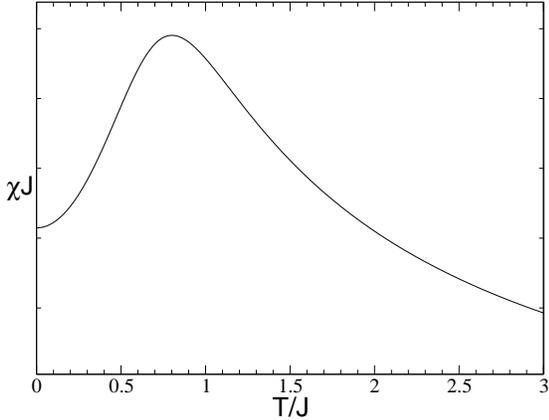}
\caption{Generic magnetic susceptibility for an antiferromagnet.}
\label{chi}
\end{figure}

Let us first consider a generic antiferromagnet, i.e.~a collection of 
spins of size $S$ which are assumed to order antiferromagnetically at low temperatures.
This implies two sublattices A and B with long range correlations throughout the lattice 
at low temperatures.  The magnetic moments are correlated in parallel on the same sublattice,
but antiparallel to the other sublattice.
The size of the order parameter may be reduced by temperature or
quantum fluctuations, but it is assumed to be non-zero.  The effect of an applied 
magnetic field $B$
can be intuitively understood as described by standard textbooks\cite{AM}.
If the field is perpendicular to the order (transverse field),  
all spins can tilt slightly towards the field
as shown in Fig.~\ref{AF-B}.  A uniform magnetization is induced in the entire sample
with a finite magnetic susceptibility 
$\chi = \partial m/\partial B = -\partial^2 F/\partial B^2$ at low temperatures.  
Note, that if the field was applied parallel to the antiferromagnetic order, 
the response would be very small,
since it costs more energy to induce a longitudinal magnetization. 
In an isotropic antiferromagnet, the order is therefore always aligned transverse to 
the field as shown in Fig.~\ref{AF-B}.  At non-zero temperatures, spin-waves become 
excited, which can be polarized with the field. Therefore the susceptibility 
increases with increasing temperature in the ordered phase.
On the other hand, 
at very high temperatures in the non-ordered phase the susceptibility is
well described by the Curie-Weiss-law $S(S+1)/3(T+\Theta)$, i.e.~decreasing with temperature.
The typical susceptibility of an antiferromagnet therefore shows a broad correlation
maximum as shown in Fig.~\ref{chi}.  Even in antiferromagnetic models which do not order 
at low temperatures this correlation maximum is well established, as for example in
the spin-1/2 chain\cite{eggert}.  In this system the entangled quantum state gives rise 
to exotic effects
such as a diverging derivative of the susceptibility with respect to $T$ as 
$T\to 0$\cite{remark}.

\begin{figure}
\includegraphics[width=0.45\textwidth]{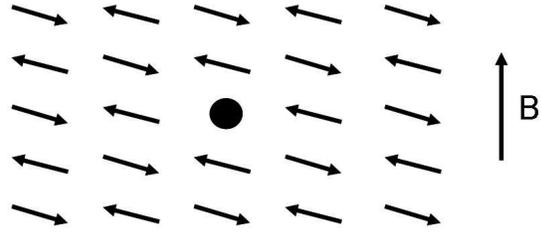}
\caption{The effect of a magnetic field $B$ on an antiferromagnet with one single
impurity (vacancy).}
\label{AF-imp-B}
\end{figure}

We will now considers what happens to the individual spins when a vacancy is 
introduced into the antiferromagnetic order on sublattice B as shown in Fig.~\ref{AF-imp-B}.
Because of the antiferromagnetic order all spins on sublattice A form one large magnetic 
moment of size $S N/2$, while all spins on sublattice B amount to a magnetic moment of $S (N/2-1)$.
The total system has therefore an effective net classical moment of size $S$ which can 
align with the magnetic field, following a Curie-law with diverging susceptibility $S^2/3T$.
This means that {\it all} spins on the sublattice A have a diverging susceptibility of
$S^2/3T$ and all spins on sublattice B have a negative diverging susceptibility of $-S^2/3T$.
This alternating Curie-type response is much larger than the uniform canting 
of the spins at low temperatures 
discussed above.  One single impurity is therefore sufficient
to induce an antiferromagnetic longitudinal order throughout the lattice 
in the presence of a magnetic field.  
So far we have neglected that in a quantum antiferromagnet the order parameter $m$
is reduced by quantum fluctuations ($m=1$ represents the maximum alignment of all spins on 
the respective sublattices).  
Since the induced longitudinal order stems 
from the existing transverse order in the system, it must also be proportional to 
$m S^2/3T$ ($m\sim 80\%$ for the spin-1/2  Heisenberg AF on a 3D cubic lattice, 
and $m\sim 60\%$ on the 2D square lattice).
Therefore, we can write approximately for the local response around an impurity, according 
to this intuitive picture
\begin{equation}
\chi(\mathbf{r}) = \chi_0  + (-1)^{r_x+r_y+1} m \frac{ S^2}{3T} , \label{chir}
\end{equation}
where $\chi_0$ is the susceptibility per site of the pure system.
The induced response is alternating (Fig.~\ref{AF-imp-B}) while the 
uniform response remains largely unchanged (Fig.~\ref{AF-B}).

We will now test this picture for the 2D spin-1/2 Heisenberg model, 
\begin{equation}
H=J\sum_{\langle i,j \rangle}\mathbf{S}_{i}\cdot \mathbf{S}_{j} \label{ham}
\end{equation}
where $\langle i,j \rangle$ denotes nearest neighbor sites on a periodic square lattice.
This model is known to exhibit 
antiferromagnetic order as  $T\to 0$.  However, for any finite temperature the
order is destroyed by quantum fluctuations.  Among the ordered antiferromagnets, this model 
is therefore at the extreme borderline to a quantum entangled state.  
The quantum Monte Carlo program we developed uses the loop algorithm in a
single cluster variety implemented in continuous
time\cite{Evertz,Evertz1,Wolff,Beard}, which gives efficient and fast updates even 
at very low temperatures.

According to Eq.~(\ref{chir}) the local response around a 
static vacancy  
\begin{equation}
\chi(\mathbf{r})=\beta\sum_i\langle S^z_i S^z_{\mathbf{r}} \rangle
\end{equation}
 can be separated into a sum of uniform and staggered parts on the lattice
\begin{equation}
\chi(\mathbf{r}) = \chi_{\rm uni}(\mathbf{r})+(-1)^{r_x+r_y}\chi_{\rm stag}(\mathbf{r}),
\end{equation}
the amplitudes of which are both slowly varying on the scale of one lattice spacing.
In order to extract those two
components we numerically extrapolate the data on the even sublattice to the odd sublattice and
vice versa and define
\begin{eqnarray}
\chi_{\rm uni}(\mathbf{r}) &=&\frac{\chi_{\rm even}(\mathbf{r})+\chi_{\rm odd}(\mathbf{r})}{2} \\
\chi_{\rm stag}(\mathbf{r}) &=&\frac{\chi_{\rm even}(\mathbf{r})-\chi_{\rm odd}(\mathbf{r})}{2}.
\end{eqnarray}

\begin{figure}
\begin{center}\includegraphics[width=.47\textwidth]{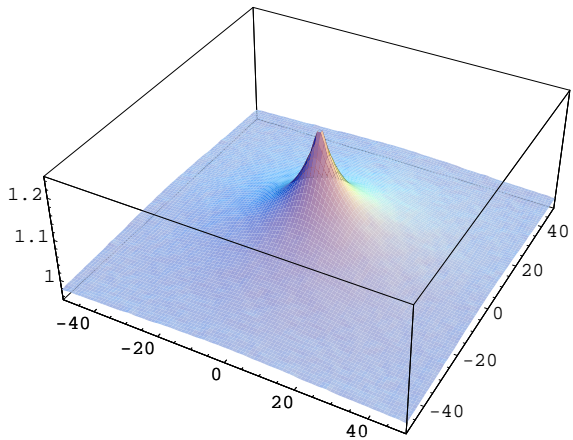}\end{center}
\caption{(Color online) $\chi_{\rm stag}(\mathbf{r})$ for T=0.05J}
\label{imp-stag}
\end{figure}
 
\begin{figure}
\begin{center}\includegraphics[width=.47\textwidth]{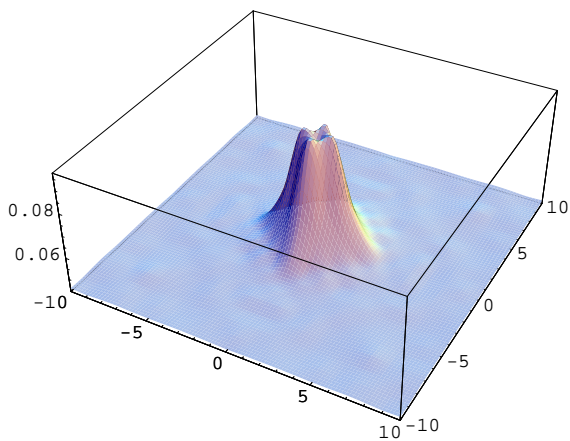}\end{center}
\caption{(Color online) $\chi_{\rm uni}(\mathbf{r})$ for T=0.05J}
\label{imp-uni}
\end{figure}

\begin{figure}
\includegraphics[width=.47\textwidth]{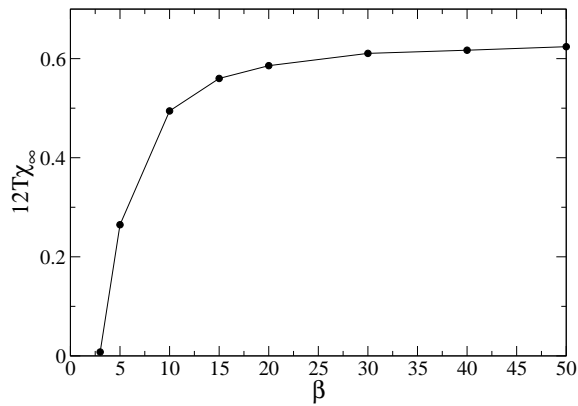}
\caption{The ratio $12 T \chi_\infty$ as a function of $\beta = 1/T$.}
\label{amplitude}
\end{figure}

The results for the staggered and uniform parts are shown in Figs.~\ref{imp-stag} and
\ref{imp-uni} for $T=0.05J$.  
The uniform part drops off very fast to the limiting value $\chi_0$, but is strongly 
enhanced around the impurity.  The staggered part also approaches a limiting value 
$\chi_\infty \sim 0.6 \frac{S^2}{3T}$, which has the expected Curie-behavior as shown in
Fig.~\ref{amplitude} in agreement with $m\sim 0.6$ for the 2D Heisenberg model\cite{reger,Sandvik2}.  
However, a broad peak around the impurity 
remains, which appears to be largely temperature independent.  Therefore Eq.~(\ref{chir})
appears to be indeed valid for longer distances from the impurity, while for shorter
distances quantum effects dominate.  We expect that this picture becomes more and more 
accurate for larger spin models in higher dimensions.  Our simulations for the 2D model
are in full agreement with previous calculations, where the impurity susceptibility 
of the entire system has been 
considered\cite{Sachdev,Nagaosa,Sandvik,Sandvik1,Sachdev2,Sachdev3,Sushkov,Sachdev1}.  For more details on the effect of more than 
one impurity, see Ref.~\cite{anfuso}.

\begin{figure}
\includegraphics[width=.47\textwidth]{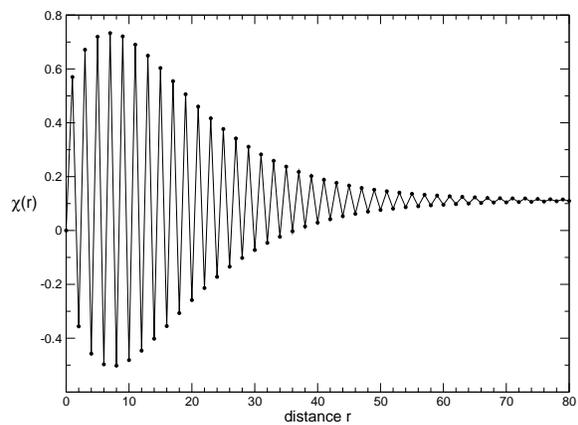}
\caption{The local response for a semi infinite chain at $T=0.05J$.}
\label{1D}
\end{figure}

We now wish to compare the situation to a model which does not order at low temperatures, 
such as the 1D spin-1/2 Heisenberg model.  In this case, vacancies cut the chain, so that we
need to consider a semi-infinite chain. This problem has been considered before\cite{eggert2}, 
with some surprising 
results.  A large alternating response is also induced by the impurity, but this {\it increases}
with the distance from the edge.  Finite temperatures, finite fields, or finite system sizes will limit 
the range of the alternating part, but generally the maximum alternating response is not closest to
the impurity site.  A typical response is shown in Fig.~\ref{1D} for $T=0.05$.  It is clear that 
such a complicated pattern is an indication of a collective state in this quantum many body system,
which gives some indication of the nature of the valence bond state\cite{dagotto}.

In conclusion we have analyzed the local response to a uniform magnetic field in 
low-dimensional spin-1/2 antiferromagnets with one vacancy.  
For the 2D model an intuitive picture of a long range antiferromagnetic
order describes the results well far away from the impurity site.  
Only the more local enhancement 
is specific to the model and must be attributed to quantum effects or corrections 
from spin-wave theory.  We therefore can argue that 
the local response is always accurately described by Eq.~(\ref{chir}) for 
larger distances from a vacancy
in {\it any} ordered antiferromagnet, where $m$ is the order parameter, which 
typically depends on spin, dimension and temperature.  Therefore a single impurity induces a 
large Curie-divergent alternating response throughout an ordered antiferromagnet, 
if a field is
applied transverse to the underlying magnetic order (this is automatically the case in an 
isotropic model).
The response in the immediate vicinity is specific to the model, but can sometimes also 
be intuitively understood, e.g.~by a valence bond basis\cite{dagotto}. 
The situation is quite different for the 1D quantum antiferromagnet, where strongly entangled
quantum states dominate the picture.  

Our results will have direct consequences on NMR and $\mu$SR experiments on doped 
antiferromagnets\cite{Mahajan,Ting,Corti,takig}.  For the one-dimensional case
the exotic boundary effects have already been confirmed, which are manifest through 
NMR satellites with a characteristic $1/\sqrt{T}$ dependence\cite{takig}.  For 
ordered antiferromagnets it is important to distinguish between the ordered
sublattice magnetization \cite{matsum} and the field induced staggered magnetization, which 
we have described here.  The field induced effects should show a more 
dramatic Curie-like temperature dependence and are assumed to be perpendicular to the 
sublattice magnetization.


\begin{thebibliography}{9}
\bibitem{remark}  A divergent slope as $T\to 0$ is surprising, because it appears to contradict
the third law of thermodynamics $\lim_{T\to 0} \Delta S \to 0$.  Therefore, 
$\frac{\partial \chi}{\partial T} =  -\frac{\partial^2}{\partial B^2} \frac{\partial F}{\partial T} 
= \frac{\partial^2 S}{\partial B^2} $ should vanish as $T\to 0$.  However, for the spin-1/2 chain the 
derivatives in respect to field and temperature do not commute in this limit.
\end{thebibliography}
\end{document}